\documentclass{article}
\usepackage{arxiv}
\usepackage[utf8]{inputenc} 
\usepackage[T1]{fontenc}    
\usepackage{hyperref}       
\usepackage{url}            
\usepackage{booktabs}       
\usepackage{amsfonts}       
\usepackage{nicefrac}       
\usepackage{microtype}      
\usepackage{lipsum}		
\usepackage{graphicx}
\usepackage[numbers]{natbib}
\usepackage{doi}
\usepackage{amsmath}

\title{Retrieving Tropical Cyclone Intensity from Climate Reanalysis using Deep Learning}

\author{
Khanh Luong \\
Department of Earth and Atmospheric Sciences\\
Indiana University Bloomington, 47405, Indiana\\
\And
Chanh Kieu\thanks{Corresponding Author: ckieu@iu.edu} \\
Department of Earth and Atmospheric Sciences\\
Indiana University Bloomington, 47405, Indiana\\
}

\renewcommand{\shorttitle}{\textit{arXiv} Deep learning tropical cyclone intensity}

\hypersetup{
pdftitle={A template for the arxiv style},
pdfsubject={q-bio.NC, q-bio.QM},
pdfauthor={David S.~Hippocampus, Elias D.~Striatum},
pdfkeywords={First keyword, Second keyword, More},
}

\begin{document}
\maketitle

\begin{abstract}
Traditional methods for improving tropical cyclone (TC) intensity from climate model outputs or projections have primarily relied on either dynamical or statistical downscaling. With recent advances in deep learning (DL) techniques, an important question is how DL can provide an alternative approach for enhancing TC intensity and structure retrieval from climate data. Using a common DL architecture based on convolutional neural networks (CNN) and a set of key environmental features relevant to TCs, we show that TC intensity and size can be effectively retrieved from climate reanalysis data without requiring super-resolution enhancement as in previous studies, even when applied to coarse-resolution climate data. This approach allows for retrieving TC intensity metrics and size that are dynamically constrained by the data, rather than estimating these quantities independently. Our results highlight that TC intensity and size are governed not only by TC internal processes but also by local environments during TC development for which DL models can learn and capture. The performance of our DL model depends on several factors such as season, the stage of TC development, or ocean basins, with root-mean-square errors ranging from $\approx$8-10 m s$^{-1}$ for the maximum 10-m wind, $\approx$10-13 hPa for minimum central pressure, and $\approx$13-21 km for the radius of the maximum wind. Although these errors are better than any direct vortex detection or statistical downscaling methods applied to the same data, their wide ranges also suggest that a 0.5$^\circ$-resolution climate data may contain limited TC information for DL models to learn from, regardless of model optimizations or architectures. Possible improvements and challenges in addressing the lack of fine-scale TC information in coarse-resolution climate reanalysis datasets are discussed.
\end{abstract}


%
%
\section{Introduction}\label{sec:introduction}
Tropical cyclone (TC) downscaling is an important procedure for improving TC representation in both climate simulations and operational forecasting \citep{HillLackman2011, Knutson_etal1998, Qian_etal2003, Tsou_etal2016, Lo_etal2008, Caron_etal2011, Bacmeister_etal2018}. While global climate or weather models are greatly valuable for capturing large-scale atmospheric circulations and long-term climate trends, they typically have coarse spatial resolutions in the range of 25–100 km even at present due to the computational limitation \citep{strandbergetal2021}. This horizontal resolution is generally insufficient to resolve TC fine-scale physical processes such as eyewall dynamics, convective bursts, or spiral bands, which control TC intensity or inner-core structures. Downscaling techniques address this gap by translating large-scale information from global models into high-resolution details, allowing more realistic representations of TC characteristics \citep[see, e.g.,][]{Xu_etal2019, Knutson_etal2007, McDonald_etal2005, Barcikowska_etal2017, Chih_etal2024, Vu_etal2024}.

From a practical perspective, typical approaches for downscaling TC intensity from climate data can be grouped into two main types. The first is statistical downscaling, which is often based on empirical or statistical relationships between some large-scale variables, such as sea surface temperature (SST), atmospheric circulation patterns, humidity, or vertical wind shear, and observed TC characteristics like frequency, track, or intensity \citep{DeMaria2010, Vecchi_etal2011, Vecchi_etal2013, Wehner_etal2010, YonekuraHall2011, Tang_etal2016, Camargo_etal2023}. Given historical data, such statistical relationships can be derived by using, e.g., regression-based models, analog analyses, or physics-based constraints, which are then applied to other global outputs to estimate future TC activities for any region or time window. 

While statistical downscaling is efficient in translating coarse-scale climate data into localized estimates of TC intensity or occurrence without the need for expensive computational resources, it assumes that empirical relationships calibrated from past observations remain valid under future climate conditions. Moreover, it cannot fully resolve physical processes governing TC structure and intensity such as the inner-core thermodynamics or small-scale convection, which are needed for capturing TC intensity or rapid intensification \cite[see, e.g.,][]{DeMaria2010, Emanuel2000, Vecchi_etal2011, Tang_etal2016, Camargo_etal2023}. 

The second type of TC downscaling, known as dynamical downscaling, uses high-resolution regional or nested models to explicitly simulate TC physical processes, thus providing physically consistent projections of TC dynamics and structure \cite[see, e.g.,][]{Denis_etal2002, Caron_etal2011, WangWu2012, Knutson_etal2013, Strachan_etal2013, Wehner_etal2017, Vu_etal2024}. 
Given lateral boundary conditions from global models, dynamical downscaling can be designed for a wide range of climate analyses and operational forecast. In particular, it provides any TC information from the model output, thus allowing in-depth analyses for all aspects of TCs. An apparent issue with dynamical downscaling is that it is sensitive to model physics, boundary conditions, model settings, and computationally expensive as compared to statistical downscaling. Thus, both downscaling methods are often used complementarily, with statistical downscaling for practical cost-effective assessments and dynamical downscaling for detailed process-based analyses.

Regardless of downscaling techniques, we note that TC intensity is mostly represented by point-like metrics such as maximum 10-m wind (VMAX), minimum central pressure (PMIN), or different variants such as accumulated energy or power dissipation index. These values are always underestimated on any climate data grid due to the fact that the actual extrema may not coincide with model grid points. One way to address this limitation is through higher-resolution dynamical downscaling \citep[e.g.,][]{Walsh_etal2007, HillLackman2011, ZarzyckiUllrich2017, Vu_etal2024, Kieu_etal2025a}. However, even with increased resolution, the inherent underestimation caused by finite grid spacing remains unavoidable. Given the strong constraints on computational resources, the large amount of climate data, and the storage associated with downscaling simulations, current TC intensity projections therefore contain significant uncertainties that one needs to further improve \cite[see, e.g.,][]{WangWu2012, Wing_etal2015, Davis2018, Sobel_etal2019, Knutson_etal2013, Vecchi_etal2019}.

Recent rapid development of machine learning (ML) techniques offers new opportunities to examine a wide range of practical problems in atmospheric and climate research \citep[e.g.,][]{murphy2012, Hastie_etal2017, Fenner2020}. In the context of short-range TC intensity retrieval and forecast, \cite{Wimmers_etal2019} presented an ML model based on a convolutional neural network (CNN)for passive microwave observations to estimate surface wind speed VMAX. By treating TC intensity as a 29-bin output of VMAX at a 5-kt increment, they obtained a promising estimation of TC intensity, although the retrieved intensity error is still relatively large ($\approx$10-14 kt). Using a different CNN architecture U-Net, \cite{liang_etal2024} showed that combining the NOAA-20 Advanced Technology Microwave Sounder data and reanalysis data could capture VMAX/PMIN effectively when verified against the best track database with RMSE of 4 kt for VMAX and $\approx$2.7 hPa for PMIN. Such a good performance in \cite{liang_etal2024} appears to be consistent with other attempts by, e.g., \cite{Hong_etal2016, Chen_etal2019, Olander_etal2021, tian_etal2023}. As discussed in \cite{Wimmers_etal2019} or \cite{Olander_etal2021}, most TC intensity retrieval models based on ML still, however, exhibit certain limitations related to, e.g., the types of intensity outputs, different ocean basins, stages of TC development, or uncertainties inherent in satellite-derived products. 

Within the context of climate downscaling, additional challenges further limit the application of ML methods to TC research. Among these, the most critical issue is the lack of sufficiently high-resolution data to train ML models capable of accurately representing TC intensity \citep{Chen_Yuan2024, Hu_etal2025, Kieu_etal2025}. One class of ML approaches specifically proposed to address this limitation is super-resolution, which can be applied in regions where very high-resolution surface wind observations are available \citep{Vandal_etal2017, Lockwood_etal2024, Chen_Yuan2024, Hu_etal2025}. While super-resolution techniques can generate finer-scale surface wind fields from which VMAX may be inferred, they are unable to estimate other intensity metrics such as PMIN or dynamical consistency among TC intensity and structure variables. Moreover, their applicability is restricted in ocean basins where high-resolution observational data are available, not to mention that the outputs of super-resolution methods remain gridded fields and thus inherit the same limitations associated with grid-point representations of TC intensity. 


Given the importance of TC intensity downscaling for practical applications, this study aims to present an ML approach for retrieving TC intensity and size from gridded climate datasets. Unlike current methods that are primarily designed for satellite or remote sensing observations for short-term weather purposes, our study focuses on deriving TC intensity and size directly from gridded climate data. This focus addresses two key objectives: (i) enhancing our ability to extract TC information from the outputs of global climate or weather models, and (ii) improving the estimation of TC intensity and structure from gridded data in support of future climate projection or analyses. With these objectives, this work can therefore be considered as the next step after performing super-resolution to derive more accurate TC intensity and size. In addition to these technical goals, our ML-based approach can also help quantify and understand the key large-scale factors governing TC intensity beyond traditional climate modeling via feature sensitivity analyses that we wish to present here as well. 

The rest of this work is organized as follows. In the next section, detailed ML model design for downscaling, training data, and experimental settings are provided. Section 3 discusses the main results, followed by the sensitivity analyses of the model performance with different model settings. Concluding remarks are given in the final section.
%
%
\section{Deep-learning designs}
\subsection{CNN architecture}
With our primary goal of retrieving TC intensity and the radius of maximum wind (RMW) from climate analysis data, the most suitable approach is to use deep learning (DL) architectures capable of processing spatial data distributions. A natural and widely used choice is CNN, which effectively detect and extract spatial features from input images. While alternative architectures such as vision transformers or diffusion models can also handle spatial data, our experiments with a few of these variants show that their performance differences in retrieving VMAX or RMW from reanalysis datasets are practically small. This is because TC structures at 0.5$^\circ$ resolution have only a limited number of features for DL models to learn. As long as TC structures and the surrounding environments influencing TC development are sufficiently distinct, CNN is sufficiently effective for this task. Therefore, we adopt CNN as the DL model of choice for TC intensity and structure retrieval in this study. 

\begin{figure}[ht]  
    \centering
    \includegraphics[height=0.25\textwidth, angle=270]{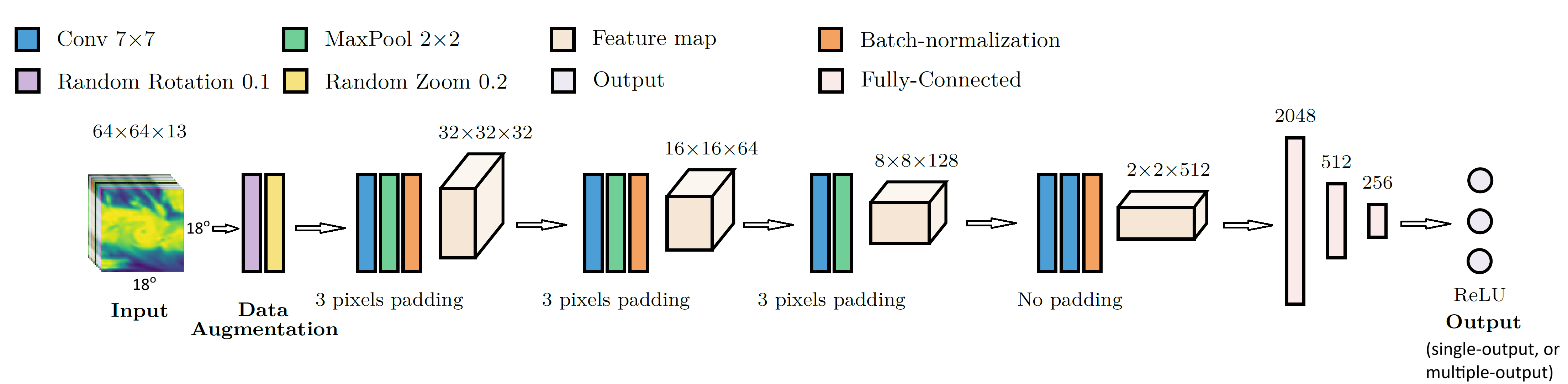}
    \caption{The default TCNN architecture for retrieiving TC intensity and size from gridded climate data, with hyperparameter information noted for each layer and corresponding operations. Note that the output from this design is either multiple-output option (all metrics VMAX, PMIN, and RMW) or single-output option (individual metric VMAX, PMIN, or RMW).}
    \label{fig:model_architecture}
\end{figure}

For this purpose, a specific CNN design for TC intensity downscaling (hereinafter referred to as TCNN model) is shown in Fig. \ref{fig:model_architecture}. This model consists of five convolutional layers, with kernel sizes of 32, 64, 128, 256, and 512, respectively. Note that our TCNN model in Fig. \ref{fig:model_architecture} is designed to process input images of $64 \times 64$ pixels with 13 channels. Each convolutional layer in the network, except for the last, employs Rectified Linear Unit (ReLU) activation, a dropout rate of 0.1, and the same padding to ensure non-linearity and maintain spatial dimensions throughout the processing stages. Note that the last convolutional layer utilizes the valid padding, aimed at reducing the output size to focus on the most relevant features. Strides are fixed at 1$\times$1 for convolutional layers and 2$\times$2 for max-pooling layers. 
The output layer consists of either a single output for each intensity metric or multiple intensity outputs simultaneously. This output layer is configured as a predictive task with \texttt{ReLU} activation, thus making it flexible for either retrieving or forecasting applications. 
The entire pipeline and model design are based on TensorFlow, which supports a variety of CNN model architectures, utilities, and visualizations.  

While this TCNN implementation is common in DL studies, we note several key points of this design that make it unique for our problem of retrieving TC intensity and size herein. First, the data augmentation step is a critical part in enhancing the performance of our model. This is because there are no known two identical TC structures, even when they have the same TC intensity. With a sample size of only $\mathcal{O}(10^4)$ for TC intensity (20 years of data, each year has $\approx$100 TCs, and each TC has a range between 50-100 cycles), these data points cannot cover the space of all possible TC shapes/structures corresponding to a given intensity. As such, introducing data augmentation with a random rotation between $\pm$45$^\circ$ not only helps enlarge the data sample but also introduces more possible structures into the model during the training. In our design, note that the augmentation is implemented once at the beginning of the training mainly to enlarge the sample size of different TC structures corresponding to a given intensity (Fig. \ref{fig:model_architecture}). While applying the augmentation during the training is generally more common, for our herein problem of TC intensity retrieval, it would make it harder to train the model in our experiments and so it is not implemented in our design.  

Second, our experiments with different designs and model hyperparameters showed that the kernel size of the CNN filter turns out to be an important factor for our problem, as it helps detect the right features from a given model resolution. For a 0.5$^\circ$ resolution data used in this study, a kernel size of 7$\times$7 is optimal for our purpose (i.e., achieving the highest accuracy training errors for all metrics). Physically, such a choice for the kernel size is not random, but it comes from the fact that TC central region has a size of $\approx$ 200 $\times$ 200 km (i.e., 4 grid points at the Modern-Era Retrospective Analysis for Research and Applications, Version 2 (MERRA-2)'s 0.5$^\circ$ resolution)
. Using a too large kernel size would smooth out TC-specific features in the storm central region, while a smaller size would introduce more noise and reduce the performance of the model.

Last, the number of CNN layers should not be too large (5 in our design) to avoid the vanishing gradient problem. One could address this issue by using a deeper network and applying some skipping mechanisms such as those used in ResNet models to overcome this issue \citep{He_etal2016, NguyenKieu2024}. However, for the current problem of extracting TC features for intensity retrieval at a horizontal resolution of 0.5$^\circ$, our experiments showed that this design suffices for extracting TC intensity without all the complications of training a deeper convolutional network. In this regard, an optimal number of CNN layers likely depends on data resolution and the retrieval problem, which requires optimizing DL models accordingly.       


\subsection{Data}
\label{subsection:data}
In this study, we used the NASA Modern-Era Retrospective Analysis for Research and Applications Version 2 (MERRA-2; \cite{merra2faq}) reanalysis data for model development and testing. MERRA-2 provides a comprehensive representation of global atmospheric structure and climate, integrating full satellite data from the post-satellite era (1980) to the present. With its global coverage and incorporation of diverse remote sensing datasets, MERRA-2 offers high-quality spatial and temporal characteristics of atmospheric parameters, making it a valuable resource for climate research and AI model training \cite{Prithvi_Schmude_etal2024}. 

Although there are several other current reanalysis datasets, we chose MERRA-2 in this study for several reasons. First, the MERRA-2 dataset provides atmospheric data in a gridded format with a horizontal resolution of $0.5^\circ \times 0.625^\circ$ in the latitudinal and longitudinal directions globally. The dataset includes all basic meteorological variables such as temperature, wind components, humidity, precipitation, or surface pressure. These variables are available at standard atmospheric levels, offering detailed vertical profiling of the atmosphere. By default, the data is provided in the Network Common Data Form (\texttt{netCDF}) format, which can be handled easily with most current ML frameworks and Python packages.

Second, most current global climate projection outputs that are given on 0.5$^\circ$ resolution. Thus, our use of 0.5-degree data can help future finetuning with other climate datasets \citep[see, e.g.,][]{Kieu_etal2025c}, thus better demonstrating its usefulness in retrieving TC intensity as expected.

One limitation of the MERRA-2 dataset as compared to other current reanalysis datasets is that this dataset contains a single resolution of 0.5$^\circ\times0.625^\circ$, while other reanalysis datasets such as European Centre for Medium-Range Weather Forecasts Reanalysis v5 (ERA5; \citealp{era5})
provides higher resolution up to 0.25$^\circ$ resolution at hourly frequency. Using such higher resolution datasets is certainly an advantage, as it can help further optimize ML models. However, whether ERA5 is better than MERRA-2 in terms of TC structure or intensity  downscaling has not been demonstrated, especially at the same 0.5$^\circ$ resolution. In this regard, our choice of MERRA-2 for this study can be considered as a pre-learning step, which can be further re-trained with ERA5 or any other datasets if needed. For the purpose of implementation and evaluation, the MERRA-2 data is therefore sufficient.  

One specific issue with the MERRA-2 dataset for TC applications is that it does not contain many meteorological variables at the surface level. For TCs with high intensity, the data at the lowest pressure level (1000 hPa) has many grid points with undefined (NaN) value \citep{merra2faq}. Feeding such incomplete data directly into any DL model would degrade its performance. To address this issue, an adaptive context-aware filling algorithm was developed, which fills all NaN values by leveraging surrounding wind field characteristics. This procedure ensures that the filled values are consistent with local atmospheric conditions while removing the NaN values that cause the non-convergence issue during training. 

For TC intensity and size labeling, the International Best Track Archive for Climate Stewardship (IBTrACS) \citep{Knapp_etal2010} compiles TC data from multiple sources into a unified, global database. This IBTrACS collects all track positions, maximum sustained wind speeds, and central pressure estimates from different meteorological agencies, thus providing comprehensive TC records of TC characteristics across ocean basins.

In this study, several key parameters including TC center locations, dates, basins, VMAX, PMIN, and RMW were extracted from IBTrACS for training and testing our DL model. These parameters VMAX, PMIN, and RMW serve as labels for our TCNN model and are paired with gridded TC information from the MERRA-2 dataset. Note that MERRA-2 provides data at fixed 6-hour intervals (0000, 0600, 1200, and 1800 UTC daily), whereas the best track dataset includes mixed 3-hour and 6-hour intervals. To ensure proper pairing, we thus omitted all IBTrACS entries that did not align with the MERRA-2 timestamps.

We should emphasize at this point that using MERRA-2 data for TC structure and then matching it with the best track intensity for training is a non-trivial problem. This is because the TC structure obtained from a 0.5$^\circ$ resolution cannot keep up with the actual intensity obtained from satellite or flight data \cite{Kieu_etal2025}. Thus, similar to the built-in assumption in \cite{NguyenKieu2024}, we will assume herein that ambient environments should contain sufficient information to determine TC intensity during the course of TC development, even with a TC structure at the 0.5$^\circ$ resolution. This is a strong assumption, as it allows us to infer TC intensity not from a given TC inner-core structure but from the environment that a TC is embedded within, to some extent the same way as the potential intensity framework provides an estimation of TC maximum potential intensity in a given environment. The rationality and justification for this assumption will be demonstrated in our model performance, which serves as a foundation for the ML approach in this study. 

\subsection{Experimental designs}
To train our TCNN model, it is first necessary to define a spatial domain that captures sufficient TC information. In this study, the default domain size is defined as an 18$^\circ\times18^\circ$ square, which is large enough to encompass both the surrounding environmental conditions and the central regions of TC circulations. Centered on each TC location reported in the IBTrACS dataset, this domain allows for the extraction of key meteorological variables from the MERRA-2 reanalysis data, which are then used as input channels for the deep learning model. While TC centers in MERRA-2 and IBTrACS can differ significantly during the early or dissipation stage of TC development, the domain size of 18$^\circ\times18^\circ$ could ensure sufficient coverage of TC environment for training a model. As a result, discrepancies in TC locations between the IBTrACS and MERRA-2 are expected to have minimal impact on our model training, even in rare cases where the center mismatch is in the range of 50–100 km. 

From an ML perspective, each domain represented by a set of meteorological variables can be treated as a multi-channel image, where each channel corresponds to a specific variable extracted from MERRA-2. For this study, thirteen MERRA-2 variables were selected to support TC intensity downscaling. These include wind components, temperature, and relative humidity at the 950, 850, and 750 hPa levels, as well as surface pressure. These channels were chosen based on our previous studies \citep{NguyenKieu2024}, which demonstrated their usefulness for TC prediction. Of course, retrieving TC intensity is a different problem  from TC formation that may require different channels. However, this difference turns out to be minor, as TC characteristics have a strong constraint among dynamics and thermodynamics, especially after they develop into a tropical depression. Thus, those channels remain useful for retrieving TC intensity and size, as will be demonstrated in this study. 


With these meteorological variables, the gridded fields were then input into the TCNN model and trained to target three observed best-track parameters including VMAX, PMIN, and RMW. Our TCNN architecture supports two configurations for retrieving TC intensity and  RMW: (i) separate models for each target variable using the same input channels (hereafter referred to as a single-output design), and (ii) a unified model that predicts all three target variables simultaneously (hereafter referred to as a multiple-output design). By comparing the model performance between these single-output and multiple-output designs, one can evaluate how strong the internal constraints among TC dynamics could govern the ability to downscale TC intensity and structure in DL models. 

With this DL design, we examined three major ocean basins including the North Atlantic (NA), the Northwestern Pacific (WP), and the Northeastern Pacific (EP), which have a total of 29,383 TC data cycles from 1980 to 2020. Of these, 29,011 cycles had corresponding MERRA-2 data available. After excluding 3,433 cases due to missing values, a total of 25,578 data points remained for further analysis. This final dataset was partitioned into training set of 39 years with 10\% for validation during training, and 1 year held out for testing and evaluation. This chronological splitting is needed to avoid overfitting issues for practical applications \cite[e.g.,][]{Chen_etal2019,NguyenKieu2024}. By further repeating this sampling process $N$ times (the so-called K-fold evaluation), we can gain a clearer understanding of how the TCNN model performs on test data containing entirely "unseen" TCs during training. 
This K-fold evaluation is essential for fully assessing the capability of our TCNN model, thereby ensuring a more robust evaluation of model performance.

For all training settings, our TCNN model was trained with 1,000 epochs and a batch size of 128. During training, input features undergo random augmentation, including a maximum rotation of 10\% and a maximum zoom of 20\%. These augmentations introduce additional variability into the dataset, thus enhancing the model's generalization. The learning rate is modulated using a sigmoid decay schedule, described by the formula:
\begin{equation}
\text{learning rate} = -0.0497 + \left(1.0 + 0.0497\right) / \left(1 + \left(\frac{\text{epoch}}{107.0}\right)^{1.35}\right),
\end{equation}
starting from an initial rate of 0.001. After reaching a minimum value of 0.0001, the learning rate is kept at this fixed value for the rest of the training.

To assess the robustness of our model, we conducted additional experiments alongside the TCNN model. Specifically, TC intensity and RMW were calculated following the storm track directly from the MERRA-2 grid using traditional vortex-tracking methods. In addition, we implemented a linear regression model using the same input channels as the TCNN. For this regression model, all input channels were spatially area-averaged to derive predictors for the regression. This setup mimics the statistical downscaling of TC intensity from the MERRA-2 dataset and serves as a reference for evaluating the extent to which the TCNN’s deep learning–based extraction of spatial information improves overall retrieval performance.

Unlike statistical downscaling approaches that estimate TC intensity metrics such as potential intensity (PI) or lifetime maximum intensity (LMI) from large-scale environmental conditions \cite[see, e.g.,][]{Sobel_etal2019, Tang_etal2016, Camargo_etal2023}, the regression model considered in this study is designed solely to infer TC intensity from low-resolution climate model output. This type of statistical downscaling, which directly maps low-resolution TC intensity to a higher-resolution estimate, is relatively uncommon in the literature \cite[see, e.g.,][]{Goswami_etal2011}, partly because high-resolution dynamical downscaling generally provides a more realistic representation of TC structure and intensity. Nevertheless, we employ this simple regression model as a baseline against which to evaluate our DL framework. That is, the regression model serves a role similar to climatology in forecast verification, providing a straightforward benchmark that helps quantify the additional skill gained by the DL approach in recovering TC intensity information from low-resolution climate data.

As a final step to verify the significance of different input channels on model performance, several sensitivity analyses were conducted by systematically removing each channel from the input dataset and then re-evaluating the model performance. Following the removal of each channel, note that the model has to be re-trained and any impacts on its performance will be based on changes in the Mean Absolute Error (MAE) and Root Mean Squared Error (RMSE) on the test set. The relative importance of each channel can be then determined by observing the degradation in the model performance, as measured by the changes in RMSE and MAE relative to the control design with full channels. In addition to assessing individual channels, we also examined the impact of a group of related channels, such as multi-level wind fields, humidity levels, or temperature fields, instead of individual channels. These comprehensive approaches allow us to quantify how an individual or combination of channels can influence the overall model's accuracy and effectiveness in retrieving TC information.

\subsection{Training metrics}
Given the nature of our TC scalar information retrieval, a common choice for the loss function and accuracy metrics is MAE, which measures the average magnitude of errors between prediction and true labels, defined as follows:
\begin{equation}
\text{MAE} = \frac{1}{N} \sum_{i=1}^N |y_i - \hat{y}_i|
\end{equation}
where \( N \) is the number of observations, \( y_i \) are the true values, and \( \hat{y}_i \) is prediction. The simplicity and direct interpretation of MAE as the average error magnitude make it suitable for monitoring our training and validation of any continuous variable. It is also robust against outliers, as the absolute values do not overly penalize larger errors, which can be crucial in datasets susceptible to noise or anomalies. 

In addition to MAE that treats all errors with equal impact on the model's adjustments, we also used the mean squared error (MSE) metric for our training. MSE is another useful metric for assessing the performance of predictive models, calculated as the average of the squares of errors as:
\begin{equation}
\text{MSE} = \frac{1}{N} \sum_{i=1}^N (y_i - \hat{y}_i)^2
\end{equation}
The squared errors in MSE indicate that larger discrepancies between predicted and true values are given greater weight, which can be advantageous in scenarios where such errors are particularly undesirable. 
Both MSE and MAE were employed as accuracy metrics in our TCNN training, allowing them to complement each other and provide a more balanced evaluation during the training process.

Along with the accuracy metrics mentioned above, we used two loss functions for our training, which include the Huber loss and the Log-cosh loss. The Huber loss function is designed to be robust to outliers, balancing the strengths of MSE and MAE \citep{Jadon_etal2022}. 
and it is given as follows:
\begin{equation}
L_{\delta}(y, \hat{y}) = 
\begin{cases} 
\frac{1}{2}(y - \hat{y})^2 & \text{if } |y - \hat{y}| \leq \delta, \\
\delta(|y - \hat{y}| - \frac{1}{2} \delta) & \text{otherwise.}
\end{cases}
\end{equation}
This loss function is more useful in datasets where the tolerance towards smaller errors is more important, while still needing to mitigate the influence of significant outliers such as RMW.

For the Log-cosh loss, it computes the logarithm of the hyperbolic cosine of prediction errors, offering a smooth curve that is always differentiable \citet{Jadon_etal2022}. This property makes it more useful for gradient-based optimization methods. The function is expressed as:
\begin{equation}
\text{LogCosh} = \sum_{i=1}^N \log(\cosh(\hat{y}_i - y_i))
\end{equation}

Throughout the training of our TCNN model in this study, both the Huber and Log-cosh loss functions were used to evaluate and select the best model in terms of MAE and MSE, using the save best model utility from Tensorflow known as \texttt{ModelCheckpoint} callback function. Results from this best model for the control workflow in Fig. \ref{fig:model_architecture} and related sensitivity analyses are presented in our next sections.

\subsection{Model accessibility}
The TCNN model is implemented in \texttt{Python} (3.10.10) and utilizes \texttt{TensorFlow} (2.18.0) with CUDA toolkit (11.7) support. The model, along with its default hyperparameters used to produce all results in the following section, is available in the Zenodo repository (https://doi.org/10.5281/zenodo.15015211). A user manual, which includes setup instructions and how to apply it to other climate datasets, is provided in the repository’s README file.

For sensitivity experiments, users are required to modify the main job script to accommodate each specific experiment. Although automating the workflow for running all sensitivity experiments is feasible, the current TCNN model release includes only a single job script \texttt{job\_control.sh} under the directory \texttt{models/TC-net-cnn}, which requires manual adjustment of hyperparameters. This deliberate design choice of manual adjustment model parameters aims to promote transparency, encourage learning, and enhance reproducibility.
%
%
\section{Results}
\subsection{Intensity retrieval benchmarking}

\begin{figure}[ht]
\centering
\includegraphics[width=0.9\textwidth]{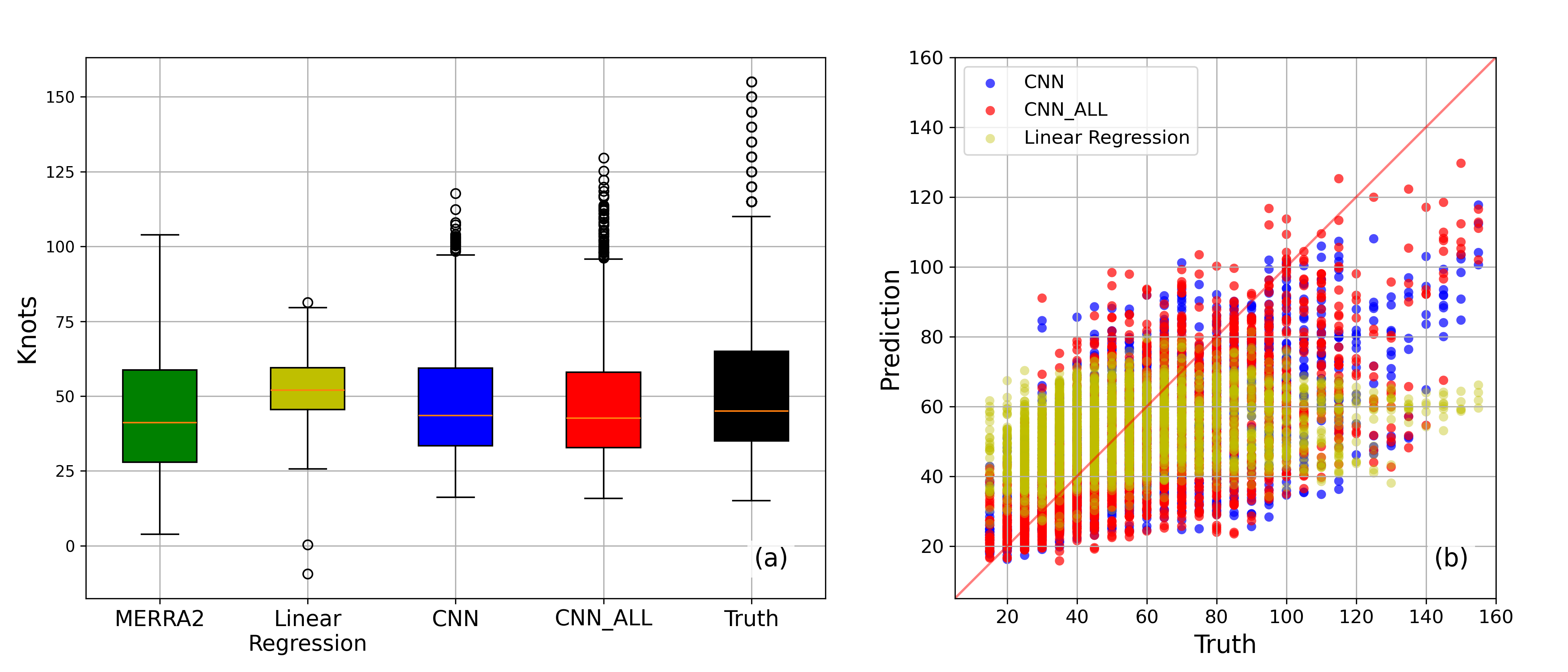}
\caption{Comparison of the predicted VMAX (unit, kt) as obtained from the TCNN model with the single-output design (blue), multiple-output design (red), directly vortex tracking on the model grid (green), statistical downscaling with a regression model (yellow), and the observed intensity from the best track (black) for the randomly-sampled test set in the form of a) box plots, and b) a scatter plot. The thin red line in (b) denotes the perfect forecast. Note that the labels CNN and CNN\_ALL in (a) denote the single-output and multiple-output design in our TCNN model, while MERRA2 denotes the results directly from the vortex tracking on the MERRA-2 grid.  }
\label{fig:wind_speeds}
\end{figure}

To evaluate the capability of our DL model in retrieving TC intensity and RMW information from gridded data, we first present results obtained from our default TCNN design shown in Fig. \ref{fig:model_architecture} for all ocean basins. For this, Fig. \ref{fig:wind_speeds} compares the distribution of TCNN-predicted VMAX with the observed intensity distribution from the test set for the single-output and multiple-output designs. As shown in Fig. \ref{fig:wind_speeds}, the TCNN model captures several key statistics of VMAX including the median, interquartile range, and the overall distribution of observed intensities with RMSE in the range of 19.1 and 19.4 kt, depending on the method of retrieving.  

Of note, both the box and scatter plots show that the multiple-output design provides a wider range of VMAX as compared to the single-output design, despite its overall underestimation of very high-intensity limit ($>$130 kts). This is important because it indicates that the TCNN model is able to capture internal constraints among different TC properties for producing this variability, using the information on the coarse-resolution grid. In contrast, the single-output design focuses only on VMAX, which has no data constraint with other TC fields. Thus, it cannot pickup other signals from pressure or structure, which are needed for the model to infer a higher VMAX limit. 

From the practical perspective, this is a non-trivial result as we recall that our training data is the MERRA-2 dataset with a resolution of 0.5$^\circ$. At this resolution, any VMAX detected directly on the gridded data or statistical downscaling would not generally match with the best track intensity (Fig. \ref{fig:wind_speeds}). In addition, the TC structure is marginally resolved by the MERRA-2 data at the high-intensity limit. As a result, retrieving VMAX on the 0.5$^\circ$ grid without account for spatial information as in the traditional vortex detection or statistical downscaling method gives worse results, with a larger RMSE of $\approx$25.9 kt and 31.5 kt as compared to the RMSE from the TCNN model (see the green and yellow box plots in Fig. \ref{fig:wind_speeds}a). In this regard, the TCNN model demonstrates that TC development leave some important spatial imprints on large-scale environments that DL models can learn from data that other traditional methods could not. To some extent, this is very analogous to the potential intensity framework, which provides the maximum intensity that a TC can obtain from a given environment with no details of TC inner-core structure or transient development \citep{emanuel1986, kieu2015, Ferrara_etal2017, KieuWang2017a, DownsKieu2020, Gilford2021}. 

\begin{figure}[ht]
\centering
\includegraphics[width=0.8\textwidth]{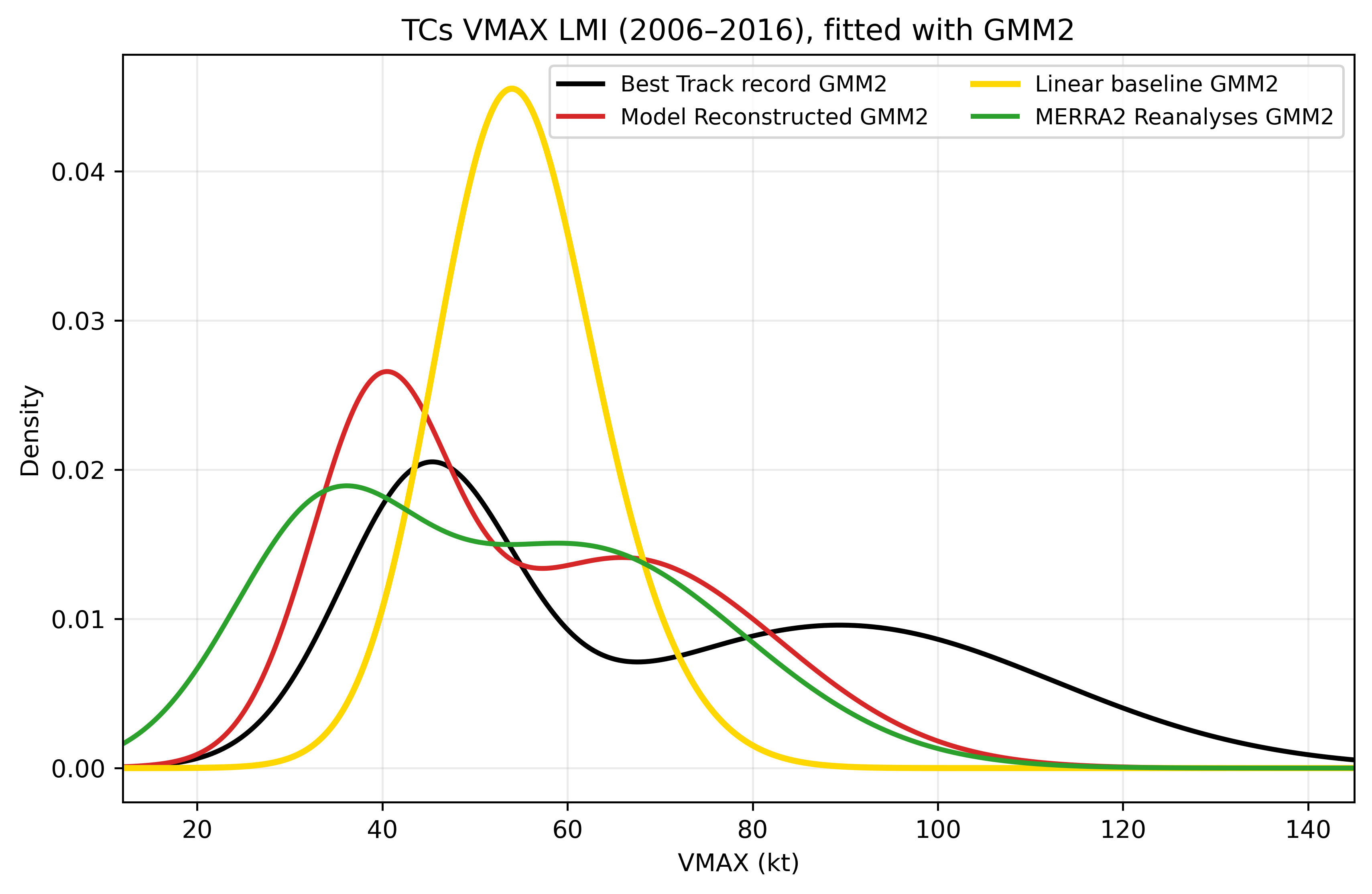}
\caption{Histogram distribution of the LMI (kt) for the test period as obtained from the multiple-output design of the TCNN model (red), direct vortex tracking (green), and best track (black).}
\label{fig:LMI}
\end{figure}

Of equal importance for verifying TCNN performance beyond the VMAX statistics are additional extreme-intensity constraints. Among these, the maximum lifetime intensity (LMI) is a common metric, as it exhibits distinctive characteristics from a climate perspective that warrant further examination \citep{Lee_etal2016, Kieu_etal2025}. Figure \ref{fig:LMI} presents the LMI distribution for the test period derived from the multi-output TCNN model, compared with best-track data and other retrieval methods.

It is evident that the vortex-tracking retrieval method exhibits a weak double-peak structure in the LMI distribution, while the TCNN substantially enhances this feature, with the second peak shifted toward higher intensities and closer to the observed distribution. In constrast, the statistical downscaling approach based on regression fails to reproduce the double-peak structure altogether, instead producing a single peak concentrated around 50–60 kt.

The improved LMI distribution as seen in Fig. \ref{fig:LMI} therefore provides additional evidence of the TCNN’s effectiveness in enhancing TC intensity estimates by incorporating ambient environmental information from coarse-resolution grids. Of course, TCNN does not fully capture the observed LMI distribution at the very high intensity range, likely due to the limited intensity-related information available at the 0.5$^\circ$ resolution for model training. Despite this limitation, the consistent improvement of TCNN in reproducing both the full VMAX and LMI distributions suggests that incorporating spatial environmental information adds significant value to TC intensity retrieval, beyond what traditional methods can achieve. 

\subsection{Dynamical and size constraints}
Given that TC intensity is characterized by several other metrics beyond VMAX, it is apparent to examine next how TCNN performs for other metrics including PMIN and RMW. Note that these metrics are not just another way to represent TC intensity and structure, but they contain some dynamical constraints among different TC dynamical constraints that we wish to further verify. Such dynamical constraints from the TCNN output are the key difference of our study as compared to previous downscaling works based on, e.g., super-resolution models.  

\begin{figure}[ht]
\centering
\includegraphics[width=0.9\textwidth]{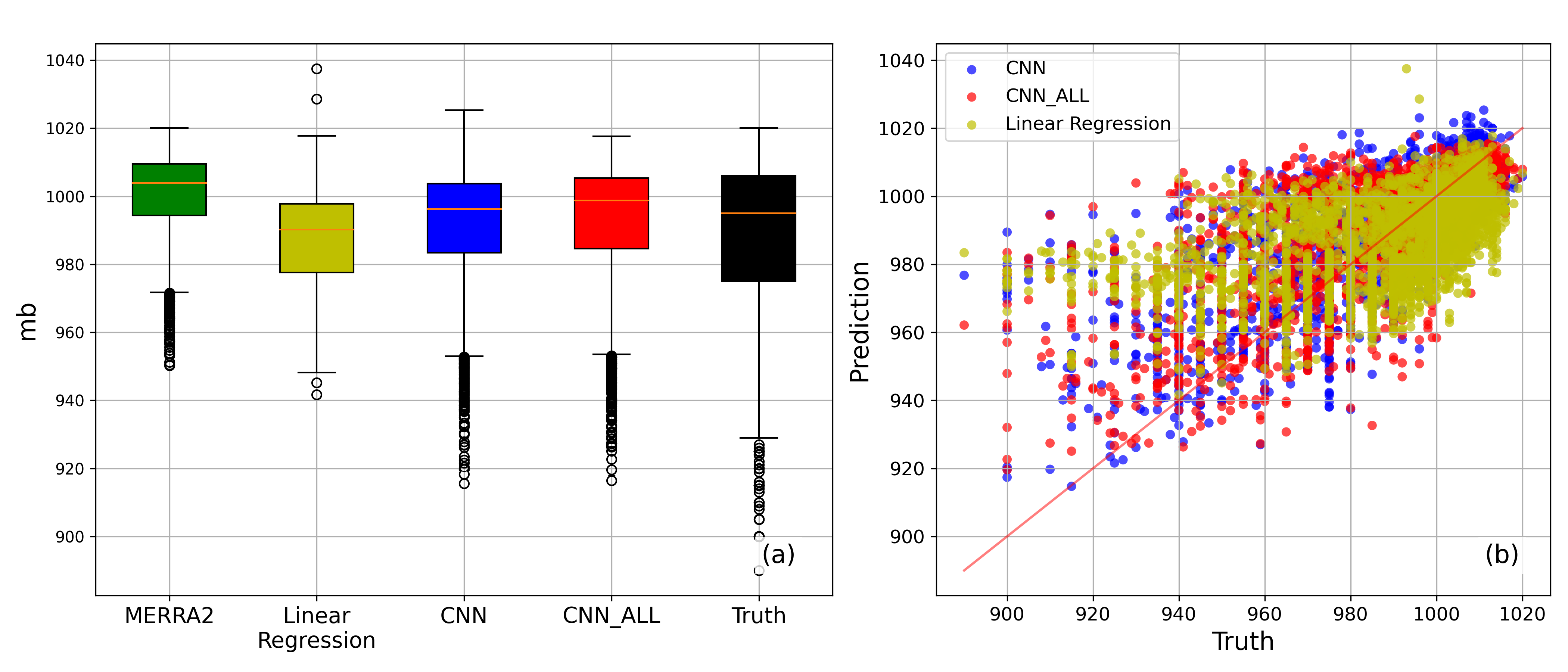}
\caption{Similar to Fig. \ref{fig:wind_speeds} but for the minimum central pressure PMIN.}
\label{fig:min_pressure}
\end{figure}

In this regard, Fig. \ref{fig:min_pressure} shows the full distribution of PMIN as obtained from TCNN model as compared to other retrieval methods and observation. Consistent with the VMAX results, the TCNN model performs well in downscaling PMIN when compared to the traditional retrieval directly from the MERRA-2 grid. For the single-output and multiple-output design, the RMSE for PMIN are $\approx$ 10.3 hPa and 10.1 hPa, respectively (Fig. \ref{fig:min_pressure}). These RMSE values are larger than those reported for in-situ observational errors \cite{Zhang_etal2019} due to the lack of detailed TC inner-core at 0.5$^\circ$ resolution, yet they are still better than the direct calculation of PMIN from the gridded data ($\approx$21.5 hPa) or statistical downscaling ($\approx$19.4 hPa). 

Unlike VMAX, we should note that PMIN in the best track data is typically a diagnosed metric, derived from an empirical pressure–wind relationship rather than direct measurements except when aircraft reconnaissance data are available \citep[see, e.g.,][]{CourtneyKnaff2009, KnaffZehr2007, Kieu_etal2010}. As a result, it is more appropriate to validate the pressure-wind relationship produced by the TCNN model against that used in the best track data rather than PMIN. 

\begin{figure}[ht]
\centering
\includegraphics[width=0.9\textwidth]{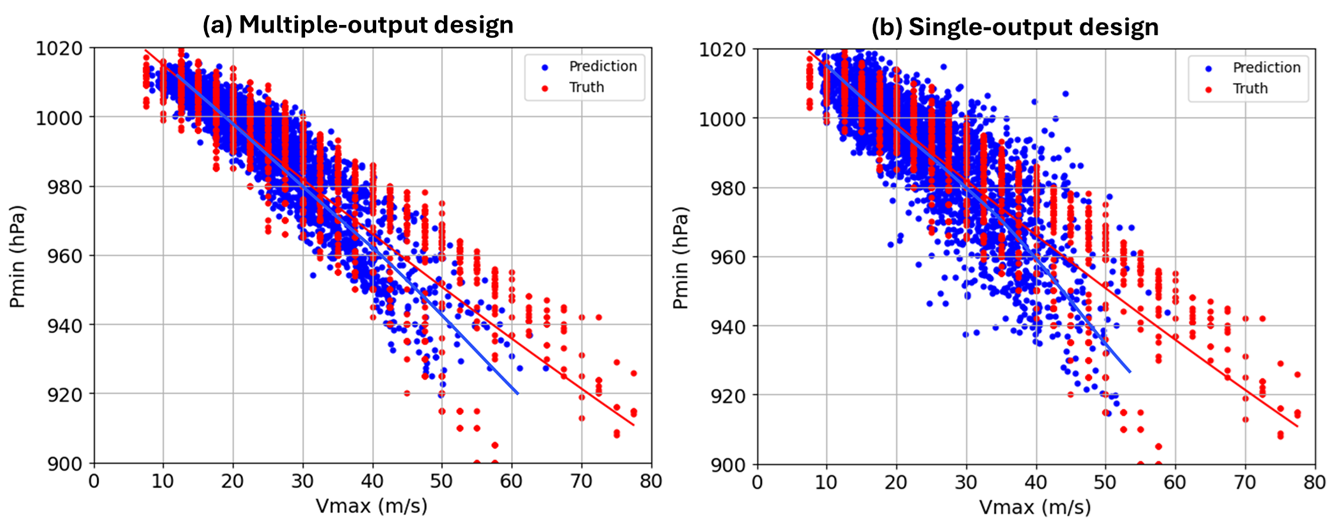}
\caption{Pressure wind relationship as obtained from the best track data (red) and that obtained from the TCNN model (blue) with (a) a multiple-output design, and (b) a single-output design. Solid thin lines denote the best quadratic fitting.}
\label{fig:pwr}
\end{figure}

Figure \ref{fig:pwr} compares the pressure-wind relationships derived from both the single-output and multiple-output TCNN designs. One can see that the multiple-output design overall provides indeed a better constraint on the model dynamics between VMAX and PMIN, as evidenced by the closer alignment of its best-fit curve toward the high-intensity tail (VMAX $>$ 50 m s$^{-1}$). In addition, the VMAX-PMIN plot shows much less scattering of data for the multiple-output design, suggesting its better constraint of the data as compared to the single-output design. However, there are still some gaps between the TCNN-derived and best track relationships in the extreme-intensity regime with VMAX $>$ 70 m s$^{-1}$ and PMIN $<$ 920 hPa, regardless of the model designs. This discrepancy at high intensity regimes is difficult to assess, because the empirical pressure–wind relationship used in the best track data is generally optimized for a broad intensity range rather than for extreme values \citep[e.g.,][]{Kieu_etal2010}. As a result, evaluating the TCNN model's performance in PMIN retrieval at these extreme intensity limits remains challenging and uncertain as seen in Figs. \ref{fig:min_pressure}–\ref{fig:pwr}.

\begin{figure}[ht]
\centering
\includegraphics[width=0.9\textwidth]{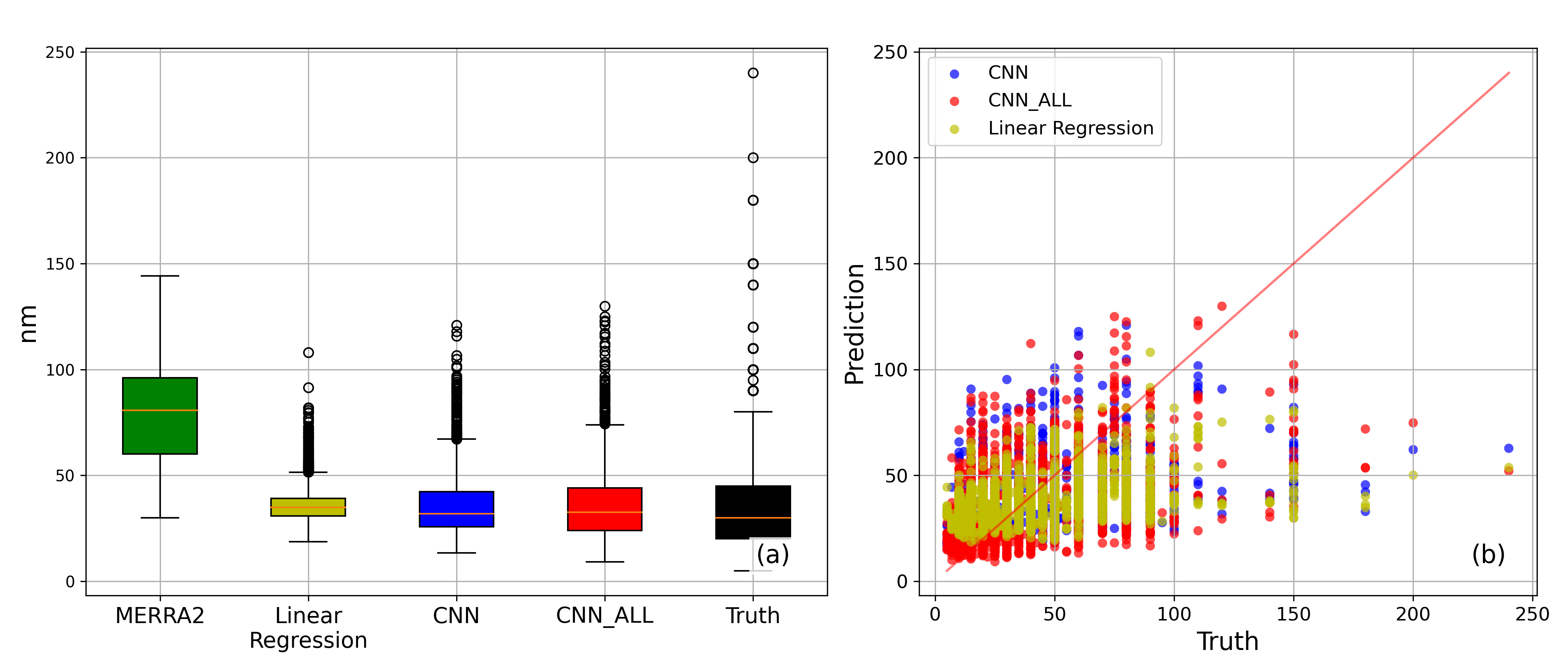}
\caption{Similar to Fig. \ref{fig:wind_speeds} but for the radius of maximum wind RMW.}
\label{fig:rmw}
\end{figure}

Along with VMAX and PMIN, our TCNN model can also provide internal TC structure information in terms of RMW as a part of its data constraints. Figure \ref{fig:rmw} presents the statistics of the RMW retrieval for both the multiple- and single-output designs. The results show that the TCNN model can reasonably capture the spectrum of RMW values up to 200 nm (320 km), with the RMSE of $\approx$22.3 nm for the TCNN model as compared to $\approx$ 23.2 and 41.7 nm for the regression and vortex tracking methods. As shown in Fig. \ref{fig:rmw}, the majority of TCs have their RMW $<$ 100 nm, for which RMW is sufficiently well-defined and consistent with TC intensity. For this regime, the TCNN model performs well in terms of mean, mode, and quartiles. Of also importance is the better performance of the multiple-output design in retrieving TCNN in terms of the range, the lower bound, as well as the upper bound than the single-output design (see the blue and red columns). This accords with the VMAX distribution (cf. Fig. \ref{fig:wind_speeds}) and suggests that internal constraints among TC intensity and RMW could help improve further the RMW retrieval for which the single-output design could not capture. 

For very large storms (RMW $>$ 250 nm), the TCNN model tends to underestimate RMW in both the single- and multi-output designs, likely due to the limited number of training samples representing such large systems. On the other hand, retrieving RMW with the vortex tracking method (green column in Fig. \ref{fig:rmw}a could provide a larger overall size, mostly because the MERRA-2 data could barely resolve the TC inner-core region, especially during the early stage of TC development. As such, computing RMW directly from the MERRA-2 grid tends to display a strong bias towards a larger TC size, thus causing a larger RMW error. 

We should mention that RMW is among the most uncertain parameters in current best-track datasets, with more consistent records only available since the satellite era. Moreover, TC size estimates are subject to various observational constraints such as satellite swath coverage, timing of observations, quadrant sampling, and environmental interference. These factors contribute to the high uncertainty of TC size, especially for storms with large RMW. Such systems are typically in weaker or less organized phases of development, making their structural characteristics more difficult to constrain based solely on VMAX and PMIN. This may explain the TCNN model’s reduced accuracy in capturing weak systems with large RMW as seen in Fig. \ref{fig:rmw}.

Regardless of the TC intensity or size metrics, the performance and evaluation of the TCNN model as presented herein are relative only to the best-track database, instead of the true (but unknown) TC intensity and structure. In fact, these best-track values for VMAX, PMIN, and RMW all contain some significant uncertainties when evaluated against direct flight data or in situ observations \citep{TornSnyder2012}. A better way to evaluate our model performance is to compute a $Z$-score that can account for both the TCNN model errors and the best-track errors, which increase the RMSE for the TCNN model \footnote{Assuming that the best track uncertainty is $\sigma_b$ and the true uncertainty $\sigma_T$ of the TCNN model. With this, the RMSE for the TCNN model with respect to the best track data is $\Gamma$ can then be estimated as $\Gamma^2 = \sigma_T^2 + \sigma_b^2$, if the best track uncertainty and the TCNN model uncertainty are independent}. Since these direct observations of TC intensity or structure are very limited at present, any DL model training with this direct observation data would not be practically useful. Thus, we have not attempted to train our TCNN model with any direct observations. After all, any observed VMAX, PMIN, or RMW from satellites or flight data still has some measurement uncertainties that one can never eliminate fully. Because of this, our evaluations of the TCNN model in this study are relative to the best-track database only.  

From a broad perspective, these results demonstrate that an optimally-tuned CNN architecture can effectively downscale TC intensity and structure from gridded climate data, significantly outperforming traditional vortex tracking or statistical downscaling methods applied to the same data. It is important to note, however, that our TCNN model has been specifically tailored to the MERRA-2 dataset at 0.5$^\circ$ resolution. Thus, applying this architecture to datasets with different spatial resolutions may require re-tuning to achieve optimal performance. Nevertheless, the current TCNN configuration remains valuable as a pre-trained model, which can be finetuned for use with other datasets of similar resolution and provides a foundation for transfer learning in similar applications. The robustness of the TCNN model across various configurations, input channels, and ocean basins will be further examined in the following sections.   


\subsection{Hyperparameter sensitivity}
Given the best-performing DL model for downscaling TC intensity and structure, it is important to next examine how the model's performance varies under different architectural or data design settings. This step is needed for assessing the robustness and applicability of our DL approach in real scenarios. While many combinations of hyperparameters and model configurations could be explored, we focus here on several key parameters that have the most significant impact on our TC intensity and structure retrieval problem, which can guide future model development with other climate datasets. 

\subsubsection{Domain size}
\begin{figure}[ht]
\centering
\includegraphics[width=0.9\textwidth]{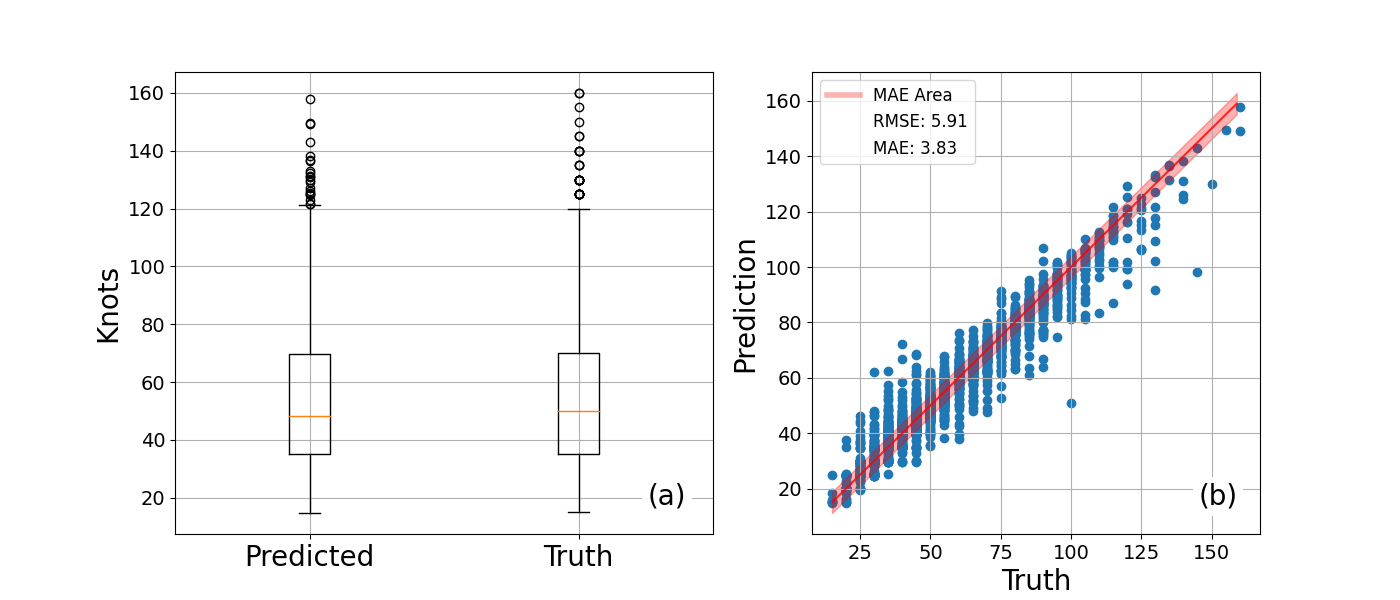}
\caption{Similar to Fig. \ref{fig:wind_speeds} but using a larger domain size of 25$^\circ \times$25$^\circ$ for the multiple-output design of the TCNN model.}
\label{fig:vmax25} 
\end{figure}

To assess first the impact of domain size on the performance of the TCNN model, Fig. \ref{fig:vmax25} presents a sensitivity analysis of this important hyperparameter. The motivation for this experiment stems from our assumption that TCs possess distinct structures and intensities influenced by their surrounding environment. Using a domain that is too small may exclude relevant environmental features, while an overly large domain could introduce irrelevant noise to the retrieval, both of which may degrade model performance, especially given the variability in TC size throughout its lifecycle. In our baseline configuration, a domain size of 18$^\circ\times$18$^\circ$ was selected, as it generally captures the key structural features of most TCs within a radius of less than 1000 km, along with the broader synoptic-scale context. In this sensitivity test, we expand the domain to 25$^\circ\times$25$^\circ$ to evaluate whether incorporating more environmental conditions improves the model's ability to predict TC intensity, particularly in cases where environmental constraints play a more significant role. For the sake of convenience, we will focus on the multiple-output design in this section, as it includes internal constraints of TC dynamics that the DL model can learn from data constraints.  

As shown in Fig. \ref{fig:vmax25}, increasing the domain size leads to a noticeable improvement in model performance, with the RMSE for VMAX decreasing from 7.1 to 5.9 knots and the MAE decreasing from 4.6 to 3.8 knots. Similar reductions are observed for PMIN (not shown), further supporting the idea that the surrounding environmental conditions play a crucial role in controlling TC intensity and structure, even when the TC inner core is not fully resolved at a 0.5$^\circ$ resolution.

Despite these improvements, using a larger domain size is not necessarily advisable for several practical reasons. First, larger domains increase the likelihood of including landmasses, which in turn reduces the sample size after NaN values are handled. As described in Section \ref{subsection:data}, our NaN-filling algorithm must be tailored to each domain size to maintain optimal model performance. Expanding the domain requires reconfiguring this process, often resulting in the exclusion of a significant portion of the training data due to land contamination. Although the smaller subset of filtered data may lead to improved performance for the $25^\circ \times 25^\circ$ domain as shown in Fig. \ref{fig:vmax25}, it limits the model’s generalizability and robustness, particularly for operational or real-time applications where a diverse and comprehensive training set is essential.

\begin{figure}[ht]
\centering
\includegraphics[width=0.9\textwidth]{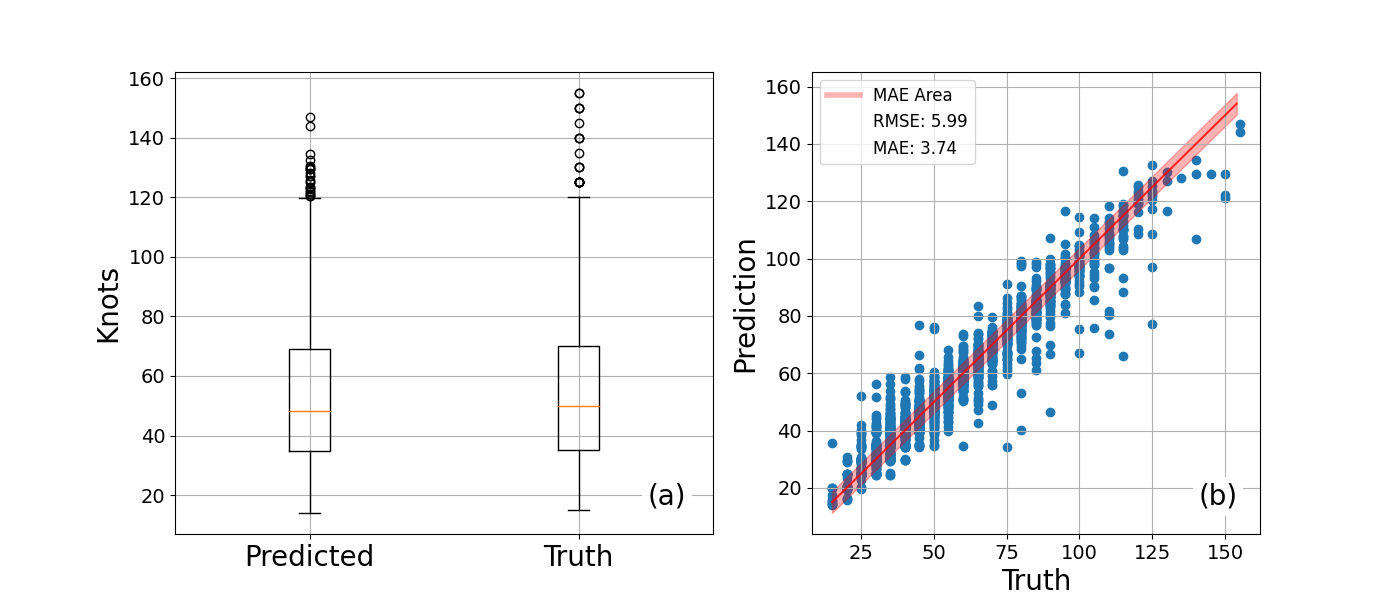}
\caption{Similar to Fig. \ref{fig:wind_speeds} but using a larger domain size of 30$^\circ \times$30$^\circ$ for the multiple-output design of the TCNN model.}
\label{fig:vmax30}
\end{figure}

Second, a larger domain might encompass more than one TC, potentially causing the model to capture unwanted TC information from nearby TCs during the active period of a TC season. The co-existence of several TCs would leave very different signals on ambient environments that DL models cannot learn due to the scarcity of those multiple-TC cases. Thus, expanding domain further would confuse DL models more. These issues with a big domain size is more apparent in our additional sensitivity experiment with a domain size of 30$^\circ\times$30$^\circ$. As seen in Fig. \ref{fig:vmax30}, such a large domain size introduces more complications to downscaling due to the external influence of far-field systems, which adversely affects TC intensity downscaling (along with an even smaller sample size as well). As a result, the model performance starts to deteriorate, with the RMSE increasing to 6.1 kts. In particular, the model becomes less effective in capturing the tail of the distribution as compared to a smaller domain size of 25$^\circ\times$25$^\circ$ or 18$^\circ\times$18$^\circ$. 

The above domain size sensitivity analyses underscore the importance of optimizing domain size for DL-based intensity and structure downscaling for a given climate dataset or global model outputs. This optimization must balance the consideration of TC-environment interactions across data resolutions while minimizing complications arising from land-sea interaction and limited sample sizes for practical applications. For the specific MERRA-2 dataset at 0.5$^\circ$ resolution, our choice of 18$^\circ\times$18$^\circ$ for the domain size is optimal and therefore chosen for all subsequent analyses. 

\subsubsection{Filter sizes and layers}
Our next set of sensitivity experiments focuses on some internal hyperparameters in the TCNN architecture. Specifically, we examine three main hyperparameters including the kernel size, the number of filters, and the number of convolutional layers. These experiments serve to justify the selected configurations for our TCNN model used in this study and provide some guidance for future development of DL models for TC downscaling. These sensitivity experiments are therefore necessary when considering that optimal hyperparameters may vary depending on each dataset used.

\begin{figure}[ht]
\centering
\includegraphics[width=0.99\textwidth]{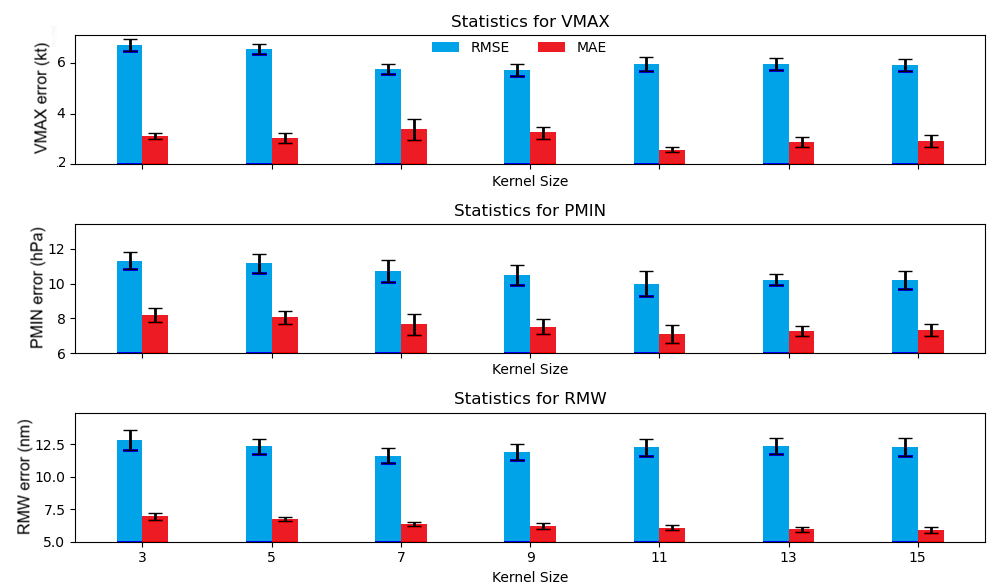}
\caption{RMSE (blue) and MAE (red) for a) VMAX (kt), b) PMIN (hPa), and c) RMW  (nm) as obtained from the validation data during the TCNN model training for a range of kernel size between 3-15 over the test dataset. Error bars denote 95\% confident intervals obtained from K-fold sampling the training and validation data.}
\label{fig:kernel-sensitivity}
\end{figure}

Fig. \ref{fig:kernel-sensitivity} shows the sensitivity of both RMSE and MAE for all three metrics VMAX, PMIN, and RMW as a function of kernel size  during the model training. One notices that the TCNN model performs best for VMAX and PMIN when the kernel size $>$ 7, and for RMW when the kernel size is between 7-9. In fact, using the single-output design for the TCNN model to predict each metric separately also shows that the model is optimal for the kernel size between 7-9. Physically, such an optimal performance of the TCNN model for kernel sizes between 7-11 can be attributed mostly to the characteristics of TC inner-core and the 0.5$^\circ$ resolution of the MERRA-2 data. Recall that the typical RMW ranges between 30-65 nm (48-100 km). Thus, a kernel size larger than 11 at a resolution of 0.5$^\circ$ would smooth out TC-specific features after several convolution and dropout operations. 

Conversely, a small kernel size would overly focus on fine details, neglecting the multi-scale relations between the TC and its ambient environment that govern TC intensity and size. In this context, the dependence of RMSE and MAE errors on kernel size shown in Fig. \ref{fig:kernel-sensitivity} is specifically tied to the MERRA-2 data and TC structure, an inherent issue when applying DL models to TC downscaling. This sensitivity justifies our choice of a kernel size of 7$\times$ 7 for the default setting of the TCNN model. 

Regarding the number of layers and filters, these two hyperparameters appear to be less important for the overall performance of the TCNN model during the model training (Table \ref{tab:sensitivity_filter_layers}). In fact, increasing the number of convolutional layers beyond 4 or using more filters does not improve model performance further, so long as the number of filters is larger than 256. This aligns with a well-known practice in CNN models, where adding more layers can lead to the vanishing gradient problem and so the model performance becomes plateaued \citep{He_etal2016}. 

While ResNet architectures with skip connections could mitigate this vanishing gradient, we stress that our input data with a resolution of 0.5$^\circ$ lacks the hierarchical features necessary to benefit from deeper CNN structures. With the limited fine-scale information in our dataset, additional layers or ResNet-type modifications are therefore unlikely to improve the model performance with more layers or filters. Unless the grid resolution is much finer ($<$3 km), we speculate that deeper architectures would offer little advantage as discussed in \cite{Kieu_etal2025}. Therefore, the issue of how many layers or filters we should use for a DL model depends on the input data, which explains why we fix our design at five convolutional layers and filter sizes of 32, 64, 128, 256, and 512 for the MERRA-2 dataset in this study as illustrated in Fig. \ref{fig:model_architecture}.

\begin{table}[h]
\centering
\caption{RMSE and MAE for VMAX as obtained from the TCNN model for different numbers of filters and CNN layers, using the validation data during the model training.}
\label{tab:sensitivity_filter_layers}
\begin{tabular}{|c|l|l|l|}
\hline
Number of CNN layers & Filter sizes & RMSE (kt) & MAE (kt) \\ 
\hline
3 & 32, 64, 64 &  15.9 & 11.1  \\ 
4 & 64, 64, 128, 128 &  15.6 & 10.7  \\
4 & 64, 96, 128, 128 &  11.9 & 7.9  \\
4 & 64, 128, 128, 256 &  9.7 & 6.5  \\
4 & 64, 128, 256, 512 &  8.6 & 5.7  \\
5 & 32, 64, 64, 128, 128 &  11.8 & 7.4  \\
5 & 32, 64, 128, 256, 512 &  7.1 & 4.6  \\
5 & 64, 96, 128, 512, 512 &  8.3 & 5.6  \\
\hline
\end{tabular}
\end{table}

\subsubsection{Basin-dependent training}
Our final sensitivity analysis examines the role of data sampling. As described in the default design and preceding sensitivity analyses, we combine TCs from all ocean basins to increase the effective training sample size. This aggregation enables the TCNN to learn from a broader range of TC morphologies and structural types associated with a given intensity. In this section, we instead focus on individual ocean basins to further evaluate TCNN performance using basin-specific TC datasets. This analysis allows us to assess how effectively the model can be trained and applied in each basin with distinct TC characteristics and large-scale environmental conditions.  

\begin{figure}[ht]
\centering
\includegraphics[width=0.5\textwidth]{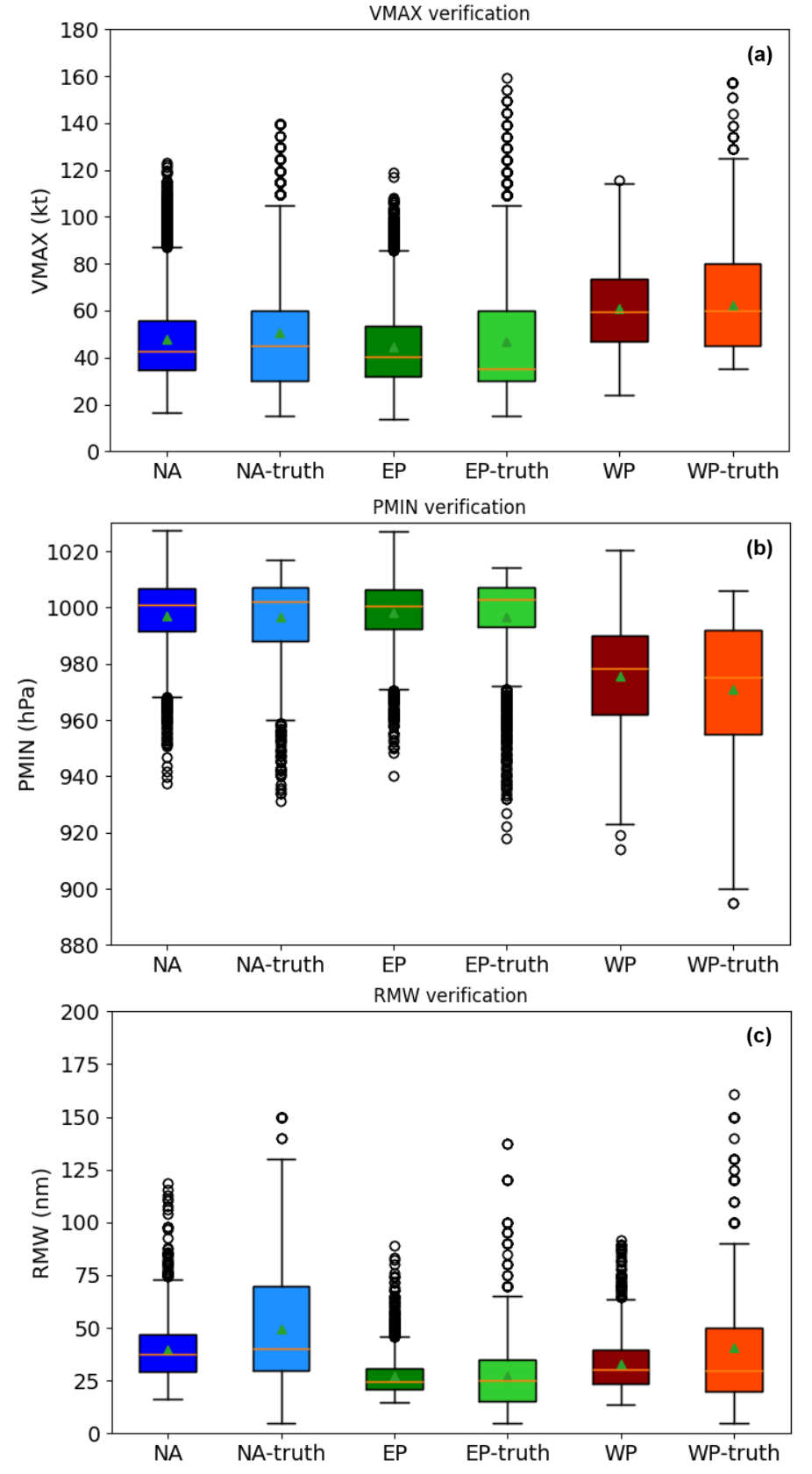}
\caption{The whisker distribution of the TCNN model prediction and the true distribution of (a) VMAX (kt), b) PMIN (hPa), and c) RMW (nm) for each ocean basin including the NA (blue), EP (green), and WP (red) basins for the test data. Red line and triangle symbol denote the median and the mean of the distribution.}
\label{fig:basin}
\end{figure}

In this regard, Fig. \ref{fig:basin} shows the full statistical distribution of the TCNN model predictions for each basin, using a one-year–leave-out sampling approach for the test data. As expected from the known TC climatology, the best-track distribution confirms that TCs in the WP basin exhibit the strongest VMAX and lowest PMIN. In addition, the EP basin displays the smallest RMW overall, as derived from the best-track data. Consistent with the overall model performance, the TCNN successfully retrieves all three metrics VMAX, PMIN, and RMW, with the best intensity performance in the NA basin, as measured by RMSE ($\approx$16.7 kt, compared to 17.2 kt and 19.4 kt in the WP and EP basins, respectively). This result is not surprising, as TCs in the NA basin tend to be weaker on average and larger in size, leading to smaller retrieval errors. In contrast, the presence of several extremely intense storms in the WP and EP basins produces outliers that increase the overall RMSE, particularly for the VMAX and PMIN metrics.

Further examination of other dynamical constraints, such as the pressure-wind relationship and lifetime maximum intensity (LMI), shows broadly similar behavior to that presented in Figs. \ref{fig:pwr} and \ref{fig:LMI}. That is, the TCNN model performs best in the WP basin, where the influence of these dynamical constraints is more evident for strong TCs (not shown). These results suggest that TCNN performance depends on whether the emphasis is placed on reproducing overall seasonal TC statistics or on capturing a small number of extreme events.

We note that this basin dependence of TCNN performance may vary with model architecture, training data, and hyperparameter choices. In fact, our additional experiments with different hyperparameter settings such as kernel size show marginal improvements for VMAX but little impact on PMIN. However, basin-specific environmental appears to have more influence on TC intensity and structure when comparing different test periods, although these effects are not statistically significant and are therefore not shown. Despite this sensitivity to input channels and configuration, the TCNN model consistently outperforms direct TC retrieval from model grids using vortex-tracking approaches or simple statistical methods.

\clearpage

The findings from these basin stratification experiments highlight an important issue that the performance of any DL model for TC intensity retrieval is sensitive to basin-specific environmental conditions. Unfortunately, addressing this issue is difficult without longer TC datasets that could cover all possible patterns of TC structures that match a given intensity in each basin \cite{Kieu_etal2025}. Given the wide range of possible TC structures for the same TC intensity, obtaining a complete TC dataset for DL model development for each basin will be much needed in the future. This limitation poses some current barriers to the practical applications of DL models to TC intensity/structural retrieval in each ocean basin, an issue that we wish to emphasize in this subsection.
%
%
\subsection{Channel importance}
While feature ranking is often treated as part of hyperparameter tuning in DL model development, we present this group of experiments separately in this subsection, as they offer more physical insights beyond merely assessing the relative importance of different channels in our TCNN model. Note that quantifying the relative contribution of each data channel to the model's performance is also a form of feature engineering that optimizes the performance of DL models, given a set of model settings and design. For the analyses presented here, we use the multiple-output design for the TCNN model, with the default kernel size of 7$\times$7 and a fixed domain size of $18^\circ \times 18^\circ$.

\begin{figure}[ht]
    \centering
    \includegraphics[width=0.8\textwidth]{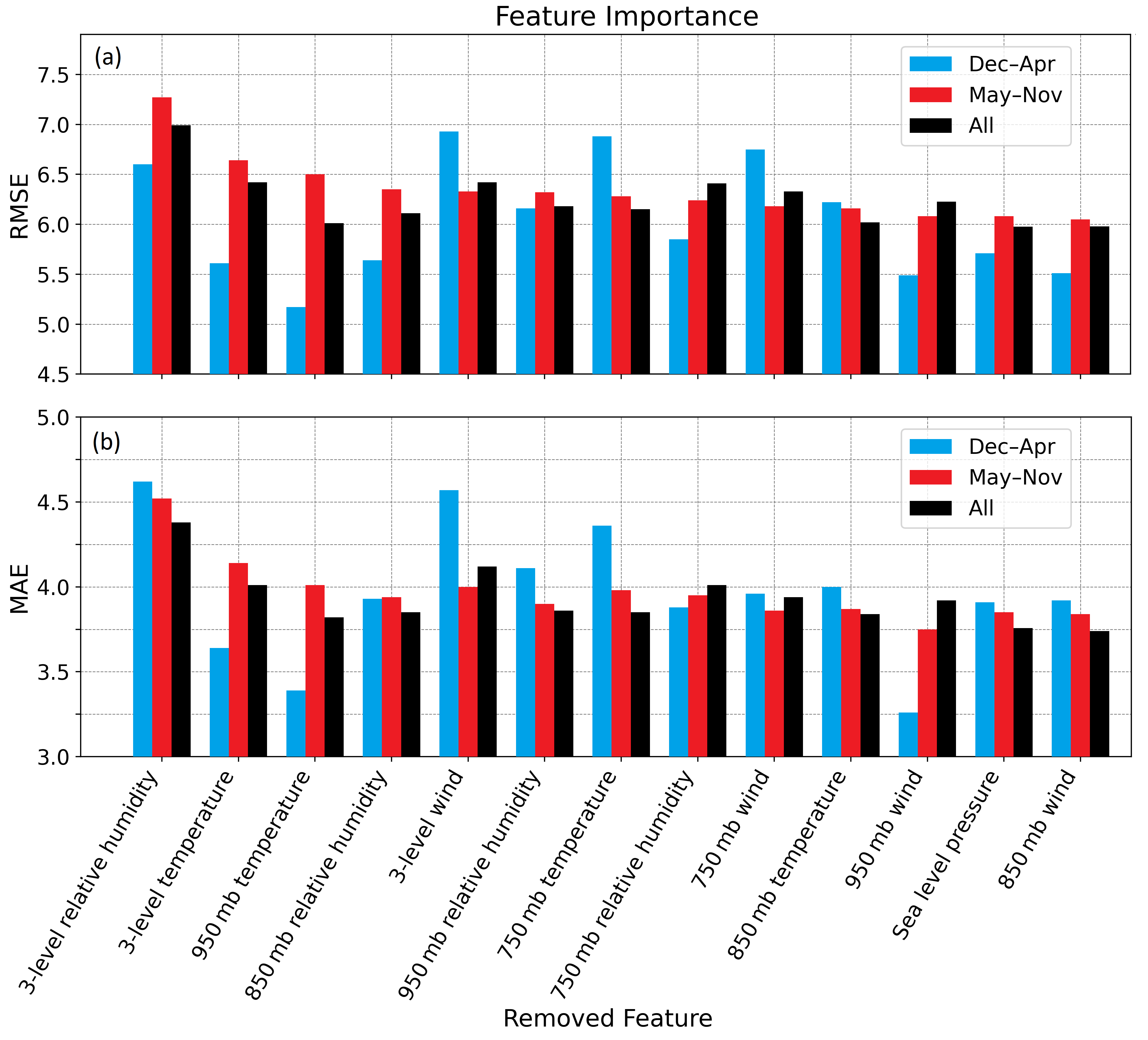}
    \caption{Bar graphs of a) RMSE (knot), and b) MAE (knot) after each input channel or a group of channels is removed during the training, using all data (black columns), data during off-peak season only (December-April, blue columns), and data during peak season only (May-November, red columns).} 
    \label{fig:channelimportance}
\end{figure}

The first notable observation from these channel ranking analyses is the impact of the moisture field on the retrieval of VMAX. As seen in Fig. \ref{fig:channelimportance} (black columns), removing individual relative humidity channels at one level 950, 850, 750 hPa, or all three levels results in the largest increase in both the RMSE for VMAX (from 7.1 to 8.2 knots) and MAE errors (from 4.6 to 5.2 knots) during the training, respectively. Consistent behaviors are also obtained for PMIN and RMW, which are expected because the TC central region tends to display a pattern of a moisture ring with a visible eye region when TCs are sufficiently strong, even on the 0.5$^\circ$-resolution grid. Such a district structure of the moisture field helps the TCNN model better recognize different development stages every time the TC moisture pattern emerges, thus contributing directly to the good performance of the TCNN model for TC intensity downscaling.    

The second behavior from this channel ranking is that the wind channels contribute inconsistently to the overall VMAX retrieval. Specifically, removing the wind channels results in a mixed behavior, with larger RMSE but somehow smaller MAE after being removed, depending on the season and level. For example, removing the 850-hPa level wind shows higher RMSE yet the MAE decreases. Likewise, removing the 950-hPa level wind reduces both RMSE and MAE significantly for the off-peak season (December-April, blue column in Fig. \ref{fig:channelimportance}). A possible reason for this inconsistency is likely due to the fact that removing 850-hPa wind may cause the TCNN model to produce abnormally higher VMAX fluctuations, which leads to larger RMSE while MAE still decreases. During the off-peak season when TCs are generally weak, the low-level wind may contain however more of the environmental winds than TC-related circulation. Thus, removing this channel helps improve the model performance during the off-peak season. For well-defined TC structures at Category 1 and above, the presence of signals from the wind channels becomes more important, thus explaining why the removal of wind channels data more impactful for the retrieval. 

On the other hand, it is intriguing to see that the sea-level pressure appears to have a smaller effect on the overall VMAX retrieval; its removal slightly increases both RMSE and MAE in our experiments. One possible reason for this could lie in the fact that sea-level pressure contains most TC intensity and size information when TCs are strong enough. For weak TCs, the sea-level pressure field at 0.5$^\circ$-resolution may not distinguish different TC strengths. So, removing this channel only changes marginally the RMSE for VMAX. Of course, this negligible effect of the sea-level pressure appears to be specific to our specific TCNN architecture and the MERRA-2 dataset. In fact, for different data stratification or larger kernel sizes, removing the sea-level pressure results in a larger increase in both MAE and RMSE errors (not shown), suggesting that sea-level pressure plays a different role in retrieving TC intensity with our TCNN model that we could not fully pinpoint.

For the rest of the channels, Fig. \ref{fig:channelimportance} indicates that removing any individual channel generally leads to an increase in RMSE. In fact, most of these channels were very similar to those found in our previous studies for TC formation \citep{NguyenKieu2024}. As confirmed in Fig. \ref{fig:channelimportance}, these same channels turn out to be also important for TC intensity retrieval, despite their varying roles between the off-peak season and the peak season. These findings emphasize the complex interplay of data channels and their impacts for different TC metrics, thus indicating the importance of proper channel selection to maximize the model's accuracy.

As one notices from the above ranking analyses, retrieving TC intensity and size depends not only on input channels but also on seasonality. That is, different months of the year should exhibit different TC characteristics, which can affect DL models. To further assess the performance of our TCNN model across seasons, we stratify the data by month and evaluate the model’s performance on these monthly-stratified subsets, using the model trained on the full dataset. This approach is chosen herein because splitting the MERRA-2 dataset into individual months for training, while preferable when ample data is available, results in a relatively small training set. By using the model trained on all data and then applying it for each month, we could ensure the robustness of our model training. 

\begin{figure}[ht]
\centering
\includegraphics[width=0.8\textwidth]{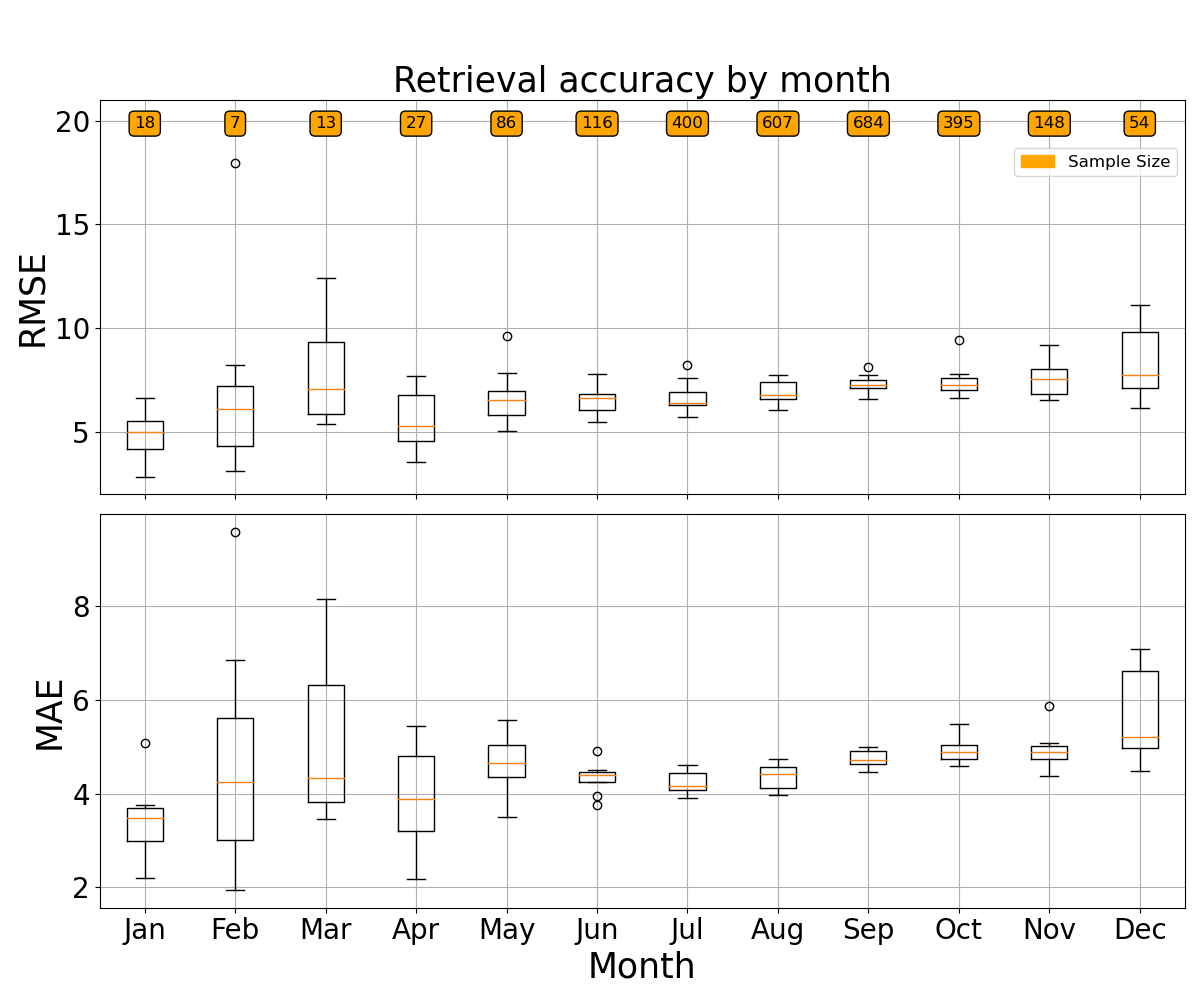}
\caption{The box plot distribution of a) RMSE (knot), and b) MAE (knot) for retrieving VMAX as obtained from our TCNN model for each month. Red lines denote the median, while the numbers in yellow boxes denote the number of TC cases that are used for intensity retrieval for each corresponding month.}
\label{fig:accuracy_m}
\end{figure}

Figure \ref{fig:accuracy_m} displays the RMSE and MAE of VMAX for different months. In general, the model demonstrates the most stable performance between June and November, which coincides with the active TC period. However, from December to May, both RMSE and MAE show larger uncertainty, with February recording the highest errors of $\approx$17.6 knots for RMSE and approximately 10 knots for MAE. Of note, the sample size from November to May is much smaller than from June to November, corresponding to a lower frequency of TCs during this period. In addition, TCs in the winter months are generally weaker, which could skew the model's performance toward weaker storm statistics. These issues together confirm higher RMSE and more variability (i.e., larger error bars) for VMAX retrieval during the winter months. In this regard, these sensitivity experiments suggest that the TCNN model performance is most stable during the peak TC season but less so during the off-peak season, consistent with the ranking analyses in Fig. \ref{fig:channelimportance}. This type of information is important for practical applications, as it allows us to understand the limitations of the TCNN model for retrieving or downscaling TC intensity and structure in future applications.
%
%
\section{Conclusions}  
In this study, we presented a deep learning (DL) approach for retrieving TC intensity and size from gridded climate data. Using NASA's MERRA-2 reanalysis dataset at 0.5$^\circ$ resolution and a CNN-based architecture, we examined a range of DL designs capable of recognizing and extracting TC information from coarse-resolution information. Evaluations of these CNN-based models (referred to as TCNN) against the best track database demonstrated that our TCNN model can retrieve TC intensity with a root mean square error (RMSE) around 16-19 kts (8-9.5 m s$^{-1}$) for VMAX and 10-11 hPa for PMIN. This performance surpasses the approach based either on vortex detection or statistical downscaling methods, given the same input data. With the 0.5$^\circ$ resolution dataset, our optimal TCNN model showed that good TC intensity retrieval can be achieved using as few as five CNN layers. The results also indicated that using a kernel size comparable to the resolved features of TCs at a given resolution, combined with sufficient nodes per layer and data augmentation tailored to the data resolution, is important to enable the model to perform effectively.

A key feature of our DL approach is its ability to simultaneously retrieve both TC intensity (VMAX, PMIN) and size (RMW). This simultaneous retrieval of intensity and structure sets it apart from conventional statistical downscaling or super-resolution DL methods, for which the estimation of VMAX is typically performed independently without PMIN or RMW. Given sufficient input data, our TCNN model can in fact learn the internal relationships between TC size and intensity metrics during training. This capability allows the model to estimate VMAX, PMIN, and RMW using a single, unified framework, thus achieving the lowest RMSE and underscoring the importance of including dynamic constraints between TC intensity and size for climate data.

Examination of the TCNN's sensitivity to different hyperparameters showed that the model's capability to retrieve TC intensity and RMW depends strongly on the nature of the input data. For the 0.5$^\circ$ resolution, the kernel size is more important when compared to the number of convolutional layers or input channels. This result suggests a way to process the input data as well as choosing proper hyperparameters for tuning DL models for each climate reanalysis dataset. Specifically, we need to ensure that the kernel size can preserve critical input information without overly smoothing it, while also avoid excessive focus on small details that may represent noise and potentially degrade the model's performance.


In addition to sensitivity to model hyperparameters, the performance of the TCNN model also depends on input channels, ocean basins, and seasonal variations. Among the various groups of input channels, the low-tropospheric moisture variables appear to be the most critical for retrieving TC intensity and size, as their removal results in the largest increase in RMSE and MAE. In contrast, horizontal winds have a relatively unclear impact on TC intensity estimation, especially during the early stage of TC development. Regarding seasonal dependence, the model's performance tends to be more reliable during the peak season as compared to the early or late TC seasons. The worse performance during the early development or peak season is due mostly to weaker intensity and fewer TCs during the off-peak season, making it more challenging for the model to achieve the same retrieval accuracy as it does during the peak season.

An important implication of our results is that even though a complete TC structure corresponding to an observed intensity is not available, DL models can still derive good information about TC intensity and size from coarse-resolution gridded data. This is in part because ambient environmental conditions do contain some key information for DL models to learn and retrieve TC intensity, even in the absence of detailed TC inner-core information. By exploring a DL model capable of learning these environmental signals, we demonstrated that DL is suitable for retrieving TC intensity and size from climate data or global model output. \textit{This is significant, as it is unlikely that we will obtain an exact TC structure corresponding to a given intensity anytime soon in the near future. Therefore, the ability to retrieve TC intensity from any gridded data without all detailed TC characteristics must rely on some detectable imprints from the surrounding environment that DL models can leverage.} 

On the other hand, our results also revealed the limitation of current climate reanalyses datasets, which are given at the resolution of 0.25$^\circ$-0.5$^\circ$. At this resolution, the TC information that one can most retrieve by DL must be limited due to the lack of fine-scale TC processes as discussed in \cite{Kieu_etal2025}, regardless of how perfect a DL model can be. How much further one can obtain TC intensity or structure from current climate datasets is an open question. Nonetheless, the results presented herein suggest that a new, different approach that can take into account the fine-scale TC processes will be needed if one wants to improve the TC intensity retrieval further. 

As a final note, we wish to emphasize that our primary goal of developing a DL model for retrieving TC intensity and structure in this study is not to achieve the best possible DL model among currently available architectures. Instead, our main objective is to demonstrate how to optimize a DL model for retrieving TC intensity and size from a coarse-resolution climate dataset, while addressing the challenges associated with different input data types, data sampling strategies, or hyperparameter selections. From this perspective, the results and approach presented in this study are informative for a proper design of DL models aimed at retrieving TC intensity and structure from global climate outputs beyond current statistical or dynamical downscaling methods. In fact, our experiments with an alternative architecture based on, e.g., the vision transformer algorithm showed slightly improved performance in terms of VMAX errors. However, the fundamental challenges such as gridded data issues, the relative importance of different input channels, different ocean basins, and the selection of model parameters tailored to a specific data resolutions are expected to remain valid and open for future research. 

\section*{Acknowledgments}
This research was funded by the NSF (AGS \# 2309929).

\section*{Author contribution} 
CK perceived the ideas, designed the workflow, analyzed the results, and wrote the draft of this work.

\bibliographystyle{unsrtnat}
\bibliography{reference_dl,reference_tcClimate,references_tc,reference_mine}  






\end{document}